\documentclass{article}

\usepackage[utf8]{inputenc}

\usepackage{graphicx}
\usepackage{color}

\usepackage[square,sort,comma,numbers]{natbib}

\usepackage{enumitem}

\usepackage{amssymb}

\usepackage{hyperref}
\newcommand{\bluehref}[2]{\href{#1}{\textcolor{blue}{#2}}}

\begin{document}

\title{Digital Scientific Notations\\ as a Human-Computer Interface\\ in Computer-Aided Research}

\author{Konrad Hinsen$^{1,2}$\\
  $^{1}$Centre de Biophys. Mol\'eculaire, CNRS;\\
  Rue Charles Sadron, 45071 Orl\'eans, France \\
  $^{2}$Synchrotron SOLEIL; L'Orme des Merisiers,\\
  91192 Gif-sur-Yvette, France
}

\date{}

\maketitle

\section*{Abstract}

Most of today's scientific research relies on computers and software not only for administrational tasks, but also for processing scientific information. Examples of such computer-aided research are the analysis of experimental data or the simulation of phenomena based on theoretical models. With the rapid increase of computational power, scientific software has integrated more and more complex scientific knowledge in a black-box fashion. As a consequence, its users do not know, and don't even have a chance of finding out, which models or assumptions their computations are based on. The black-box nature of scientific software has thereby become a major cause of mistakes. The present work starts with an analysis of this situation from the point of view of human-computer interaction in scientific research. It identifies the key role of digital scientific notations at the human-computer interface, and describes a proof-of-concept implementation of such a digital scientific notation for scientific models formulated as mathematical equations.

\section{Introduction}

Computers have profoundly changed the way scientific research is done. While the same statement can be made about many other human activities, the impact of computers on scientific research goes beyond their use as mere tools for managing data, writing articles, or communicating with colleagues. Computers process information, and information is the core resource of science. In the natural sciences, practically all results that are obtained from experimental or theoretical work are at some point processed by computers. Algorithms and their implementations in software have become an integral part of the models and methods that scientists apply and study.

The first few decades of computer-aided research have been marked by the development and application of new computational techniques. They have permitted the exploration of ever more complex systems with ever better precision, but also lead to completely new styles of scientific enquiry based on analyzing large amounts of data by statistical methods. However, the initial enthusiasm about the new possibilities offered by computer-aided research has been dampened in recent years as scientists began to realize that the new technology also brings new kinds of problems. Errors in software, or in the way software is applied, are the most obvious one \citep{MeraliComputationalscienceError2010,SoergelRampantsoftwareerrors2014}. A more subtle problem is the widespread non-reproducibility of computational results, in spite of the fact that computations are fully deterministic \citep{ClaerboutElectronicdocumentsgive1992,StoddenEnhancingreproducibilitycomputational2016}. But perhaps the most insidious effect of the use of computers is that scientists are losing control over their models and methods, which are increasingly absorbed by software and thereby opacified, to the point of disappearing from scientific discourse \citep{HinsenComputationalscienceshifting2014}. As I will discuss in section~\ref{sec:scientific-discourse}, the consequence is that automated computations are often no longer verifiable, which is an important cause of errors in computer-aided research.

In the philosophy of science, these practical questions and more fundamental ones that practitioners do not tend to worry about, are discussed in the context of the \textit{epistemic opacity} of automated computation \citep{ImbertComputerSimulationsComputational2017}. The overarching issue is that performing a computation by hand, step by step, on concrete data, yields a level of understanding and awareness of potential pitfalls that cannot be achieved by reasoning more abstractly about algorithms. As one moves up the ladder of abstraction from manual computation via writing code from scratch, writing code that relies on libraries, and running code written by others, to having code run by a graduate student, more and more aspects of the computation fade from a researcher's attention. While a certain level of epistemic opacity is inevitable if we want to delegate computations to a machine, there are also many sources of accidental epistemic opacity that can and should be eliminated in order to make scientific results as understandable as possible.

As an example, a major cause for non-reproducibility is the habit of treating an executable computer program, such as the Firefox navigator or the Python interpreter, as an abstraction that is referred to by a name. In reality, what is launched by clicking on the Firefox icon, or by typing ``python'' on a command line, is a complex assembly of software building blocks, each of which is a snapshot of a continuous line of development. Moreover, a complete computation producing a result shown in a paper typically requires launching many such programs. The complexity of scientific software stacks makes them difficult to document and archive. Moreover, recreating such a software stack identically at a later time is made difficult by the fast pace of change in computing technology and by lack of tool support. A big part of the efforts of the Reproducible Research movement consists of taking a step down on the abstraction ladder. Whereas the individual building blocks of software assemblies, as well as the blueprints for putting them together, were treated as an irrelevant technical detail in the past, this information is now realized as important for reproducibility. In order to make it accessible and exploitable, many support tools for managing software assemblies are currently being developed.

The problem of scientists losing control over their models and methods, leading to the non-verifiability of computations, has a similar root cause as non-reproducibility. Again the fundamental issue is treating a computer program as an abstraction, overlooking the large number of models, methods, and approximations that it implements, and whose suitability for the specific application context of the computation needs to be verified by human experts. To achieve reproducibility, we need to recover control over what software we are running precisely. We must describe our software assemblies in a way that allows our peers to \textit{use} them on their own computers but also to \textit{inspect} how they were built, for example to check if a bug detected in a building block affects a given published result or not. To achieve verifiability, we need to recover control over which models and methods the software applies. We must describe our model and method assemblies in a way that allows our peers to \textit{apply} them using their own software but also to \textit{inspect} them in order to verify that we made a judicious choice. Reproducibility is about the \textit{technical} decomposition of a computation into software building block. Verifiability is about the \textit{scientific} decomposition of a computation into models and methods. As I will show in section~\ref{sec:example}, these two decompositions do not coincide because they are organized according to different criteria.

In this article, I take the point of view that accidental epistemic opacity should be treated as an issue of human-computer interaction in computer-aided research. Since the problem of non-reproducibility is relatively well understood by now, even though effective solutions remain to be developed for many situations, I will focus on non-verifiability as the major unresolved problem, and in particular on the role of digital scientific notations.

\section{Motivation}
\label{sec:motivation}

The topics I will cover in this article will probably seem rather abstract and theoretical to most practitioners of computer-aided research. The two personal anecdotes in this section should provide a more down-to-earth motivation for the analysis that follows. Readers who do not need further motivation can safely skip this section.

In 1997, I wrote an implementation of the popular AMBER force field for biomolecular simulations \citep{Cieplaksecondgenerationforce1996} as part of a Python library that I published later \citep{Hinsenmolecularmodelingtoolkit2000}. A force field is a function $U(X, \Phi, G)$ expressing the potential energy of a molecular system in terms of the positions of the atoms, $X$, a set of parameters, $\Phi$, and a labelled graph $G$ that has the atoms as vertices, the covalent bonds as edges, and an ``atom type'' label on each vertex that describes the chemical environment of the atom. Force fields are the main ingredients to the models used in biomolecular simulation, and the subject of much research activity, leading to frequent updates. The computation of a force field involves non-trivial graph traversal algorithms that are habitually not documented, and in fact hardly even mentioned, in the accompanying journal article, which concentrates on describing how the parameter set $\Phi$ was determined and how well the force field reproduces experimental data. I quickly realized that the publication mentioned above plus the publicly available parameter files containing $\Phi$ with their brief documentation were not sufficient to re-implement the AMBER force field, so I started gathering complementary information by doing test calculations with other software implementing AMBER, and by asking questions on various mailing lists.

One of the features of AMBER that I discovered came as a surprise: its potential energy function depends on the order in which the atoms were listed in the configuration file that defines the initial configuration of the molecules. This makes no sense at all: interaction energies depend on the nature of the system, describe by the graph $G$ and the parameters $\Phi$, and on its instantaneous configuration $X$, but not on how the system is represented in a specific file format. I can only speculate about the cause of this design decision, but it is probably the result of simplicity of implementation taking priority over physical reasonableness, and the decision might well have been taken by an inexperienced graduate student. A reviewer of the paper would surely have objected had the feature been described there. However, the feature wasn't documented anywhere else than in the source code of a piece of software, which was never peer reviewed at all.

Over the years, I have mentioned this feature to many colleagues, who were all as surprised as I was, and often believed me only after checking for themselves. To the best of my knowledge, no paper and no software documentation mentions this behavior. I can think of only three ways to stumble on it: by experimenting with the atom order in input files, by reading the source code of a simulation program, or by talking to someone who happens to know. In fact, I am not even sure that all software implementing AMBER handles this feature in the same way, given that is in general impossible to obtain identical numbers from different software packages for many other reasons.

Pragmatists might ask how important this effect is. I don't think anyone can answer this question in general. The numerical impact on a single energy evaluation is very small. But Molecular Dynamics is chaotic, meaning that small differences can be strongly amplified. There are examples of changes assumed to be without effect on the results of MD simulations turning out to be important in the end (e.g. \citet{ReisserRealCostSpeed2017}). The hypothesis that AMBER's atom order dependence has no practical importance would have to be checked for all possible applications of the force field. It would clearly be less effort for everybody to simply fix the force field definition.

Nearly twenty years later, I was working on the gas-phase structure of small peptides. A nice study combining experiments and simulation, on peptides similar to those I was interested in, caught my interest \citep{JarroldHelicesSheetsvacuo2007}. I decided to re-do the simulations, using my own software, as a warm-up exercise. But I found very different structures. I must have done something differently... so let's find out what! Unfortunately, I didn't get very far. The details about the simulations that are given in this paper are above average for the field, and the simulation software was published. And yet, the available information was completely insufficient for understanding what the author really did. Ten years after publication, the precise software version used was no longer available. The description of the force field parameters didn't match the files I could find in a later release. As for the simulation protocol, it didn't help me much to learn that ``in some cases ... more sophisticated methods were used'', with no more detail given.

Readers familiar with the recent efforts to improve computational reproducibility might say that all that's missing is a complete archive of the author's software and input files. But I doubt that would be true. Comparing my calculations with the original author's work by going through hand-optimized source code and machine-specific workflows might be possible in principle, but not in practice. The two main ingredients to these computations are (a) a force field and (b) a heuristic search algorithm for a global minimum. Extracting either one from optimized source code is as hopeless a task as extracting high-level source code from binary executables. Comparing them, and swapping ingredients (e.g. my force field plus his minimization algorithm) are not realistic endeavors at this time.

The long-term goal of the research described in this article is to be able to express complex scientific data such as force fields and minimization algorithms in a way that permits their inspection and verification by human readers. Force fields should be published in a form that gives peer reviewers a chance to detect unphysical features, and users to perform comparisons with other force fields. Verifiability, like reproducibility, should become a requirement for computations in order to ensure the transparency of scientific research. But as a first step, it must become a possibility.

\section{Verification and validation in science}
\label{sec:verification}

The overall goal of science is to acquire reliable knowledge about the world around us. A major obstacle to achieving reliability is the unreliability of the individual scientist. Like all humans, scientists make mistakes, and the pursuit of non-scientific goals such as wealth or social status often interferes with the quest for knowledge. The scientific community has therefore established an error-correction protocol whose main ingredients are peer verification and continuous validation against new observations.

Peer verification consists of scientists inspecting their colleagues' work with a critical attitude, watching out for mistakes or unjustified conclusions. In today's practice, the first round of critical inspection is peer review of articles submitted to a journal or conference. Peer review tends to be shallow, as few reviewers re-do experiments or computations. But peer verification does not stop after publication. If a contribution is judged sufficiently important, it will undergo continued critical inspection by other scientists interested in building on its results.

Validation against new observations is a much slower process. It comes down to continuously checking the coherence of all observations and models in a given domain of science. When contradictions appear, additional experimental or theoretical work is required to figure out what went wrong. The outcome can be a minor correction such as ``observation X turned out to be faulty'', or a major correction such as ``classical mechanics is insufficient to explain the behavior of matter at the atomic scale''.

Descriptions of the scientific method tend to emphasize the role of validation as the main error correction technique. It is true that validation alone would in principle be sufficient to detect mistakes. However, the error correction process would be extremely inefficient without the much faster verification steps. A missing factor 2 in a calculation can be found more rapidly and more reliably by re-doing the calculation than by constructing an apparatus to validate the result experimentally. Verification gains in importance as science moves on to more complex systems, for which the total set of observations and models is much larger and coherence is much more difficult to achieve. As an illustration, consider a bug in the implementation of a sequence alignment algorithm in genomics research. Sequence alignment is not directly observable, it is merely a first step in processing raw genomics data in order to draw conclusions that might then be amenable to validation. The path from raw data to the prediction of an observable quantity is so long that finding the cause of a disagreement would be impossible if the individual steps could not be verified.

Verification is possible only if individual scientific contributions are sufficiently complete that a competent reader can follow the overall reasoning and evaluate the reliability of each piece of evidence that is presented. For computer-aided research, this has become a major challenge. A minimal condition for verifying computations is that the software is available for inspection, as K.V.~Roberts called for as early as 1969 in the first issue of the journal \textit{Computer Physics Communication} \citep{Robertspublicationscientificfortran1969}. His advice was not heeded: most scientific software was not published at all, and sometimes even thrown away by its authors at the end of a study. Many widely used software packages were distributed only as executable binaries, with the explicit intention of preventing its users from understanding their inner workings. This development has by now been widely recognized as a mistake and the Reproducible Research movement has been making good progress in establishing best practices to make computations in science inspectable \citep{StoddenEnhancingreproducibilitycomputational2016}.

However, the availability of inspectable source code is only the first step in making verification possible. Actually performing this verification is a complicated process in itself, which is often again subdivided into a verification and a validation phase. In the context of software, verification is usually defined as checking that the software conforms to its specification, whereas validation means checking that the specification corresponds to the initial list of requirements. Since the nature of requirements and specifications varies considerably between different domains of application, there is no consensus about the exact borderline between verification and validation. However, as for the scientific method, the general idea is that verification is a faster and more rigorous procedure that focuses on formal aspects, with subsequent validation examining how the software fits into its application context.

\begin{figure}[ht]
\centering
\includegraphics[width=8cm]{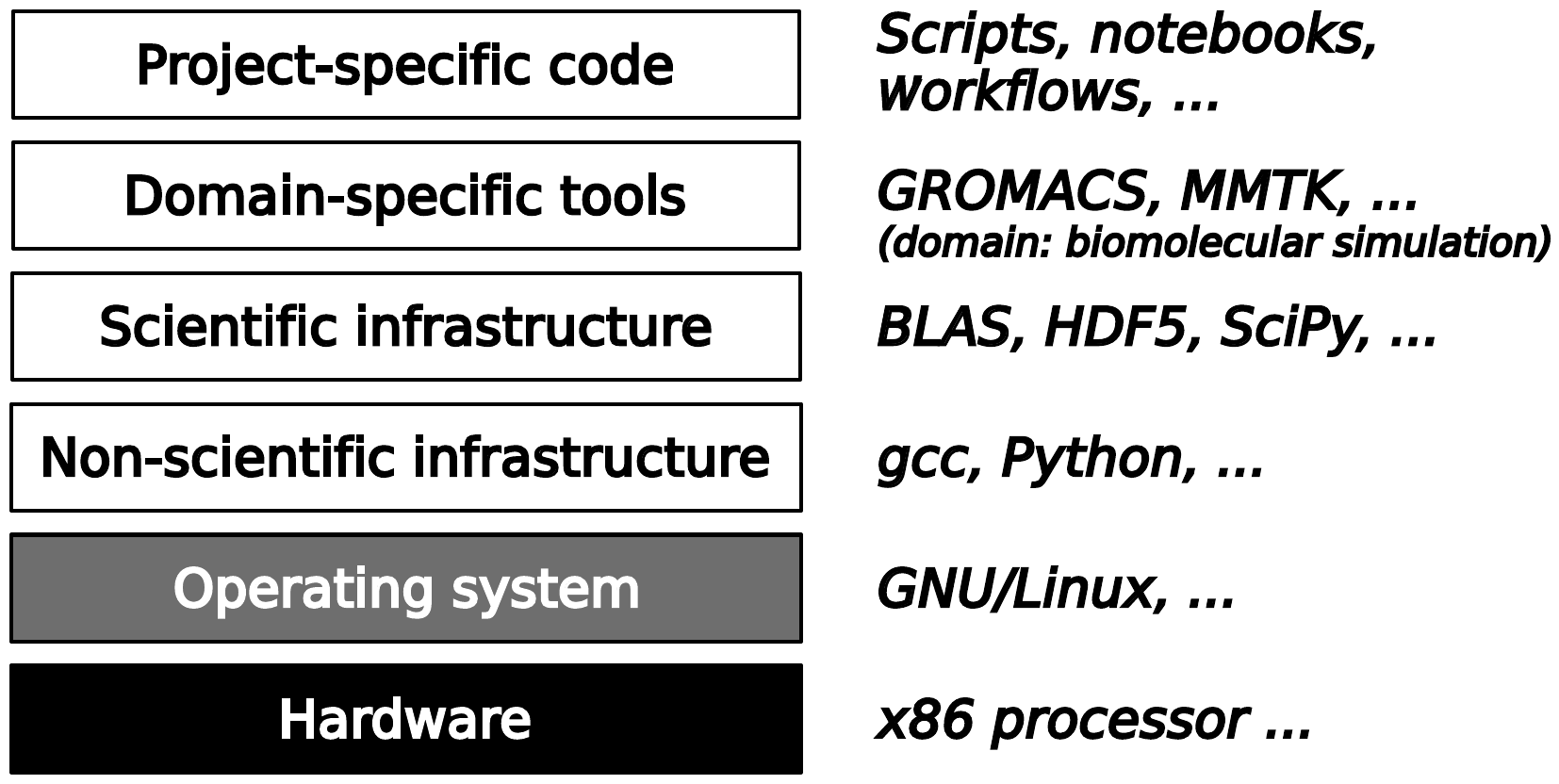}
\caption{A typical software stack in scientific computing consists of fours layers on top of hardware and systems software. The lower two layers contain widely used infrastructure software that can be verified using generic techniques from software engineering. The upper two layers are specified by scientific discourse and must be verified in its context.}
\label{fig:software-stack}
\end{figure}

Today's practice concerning verification and validation of scientific software varies considerably between scientific disciplines. Well-established models and methods, widely used software packages, and direct economic or societal relevance of results are factors that favor the use of verification and validation. Independently of these domain-specific factors, the place of a piece of software in the full software stack required for a computation determines which verification and validation techniques are available and appropriate. The typical four-layer structure of this software stack is shown in Fig.~\ref{fig:software-stack}. On a foundation consisting of hardware and systems software, the first layer consists of infrastructure that is not specific to scientific research, such as compilers. From the scientist's point of view, these are commodities provided by the outside world. The next layer consists of scientific infrastructure software that provides widely used algorithms, e.g. for numerical analysis or data management. This generally stable software is developed for research and in contact with scientists, but in terms of verification and validation can be handled like non-scientific infrastructure, because the scientific knowledge embedded into this software consists only of well-known models and methods and the software has a clear though typically informal specification.

The upper two layers are the most difficult ones to verify because there their specifications are incomplete or non-existent. The top layer consists of code written for a specific research project, with the goal of computing something that has never been computed before. It is also unlikely to be reused without modification. This makes it impossible to apply standard testing procedures. Moreover, quite often this top layer consists of scripts that launch executables written independently and in distinct programming languages, making it difficult to exploit language-centric verification approaches such as static type checking.

One level below, there is a more stable layer of domain-specific tools, which are developed by and for communities of scientists whose sizes range from one to a hundred research groups. In fundamental research, where models and methods evolve rapidly, this domain-specific software is almost as difficult to verify as the project-specific layer. Moreover, it is typically developed by scientists with little or no training in software engineering. For many years, verification and validation of domain-specific tools was very uncommon in fundamental research. Today, widely used community-supported software packages use quality assurance techniques such as unit testing and sometimes continuous integration. However, the scientific validity of the software is still not systematically evaluated, and it is not even clear how that could be achieved. Journals dedicated to the publication of scientific software such as the \bluehref{https://joss.theoj.org/}{Journal of Open Source Software} \citep{SmithJournalOpenSource2017} or the \bluehref{https://openresearchsoftware.metajnl.com/}{Journal of Open Research Software} do not even ask reviewers to comment on scientific correctness because such a request would be unreasonable given the current state of the art.

The basic difficulty with verifying and validating the top layers of scientific software is the lack of clear specifications. The core of such a specification would be made up of the models and methods that are applied. They are, however, exactly what researchers modify in the course of their work. As a consequence, each computation requires its own \textit{ad-hoc} specification that combines some established models and methods with some more experimental ones into a whole that is usually too complex to be written down in a useful way. The closest approximation to an informal specification is the journal article that describes the scientific work. An essential part of verification is therefore to check if the computation correctly implements the informal description given in the article, or inversely if the journal article correctly describes what the software does. To understand the challenges of this step, it is useful to take a closer look at the interface between scientific discourse and scientific software.

\section{Informal and formal reasoning in scientific discourse}
\label{sec:scientific-discourse}

The main purpose of scientific discourse, whose principal elements today are journal articles and conference presentations, is to communicate new findings in a way that permits peer verification and re-use of the findings in later research. Another category of scientific discourse serves pedagogical purposes: review articles and monographs summarize the state of the art and textbooks teach established scientific knowledge to future generations of scientists. A common aspect of all these writings aimed at experts or future experts is an alternation of informal and formal reasoning. More precisely, formal deductions are \textit{embedded} in an informal narrative.

Before the advent of computers, formal deductions were mainly mathematical derivations and occasional applications of formal logic. In this context, the formal manipulations are performed by the same people who write the informal narratives, with the consequence that the back and forth transitions between the two modes of reasoning, informal and formal, is often blurred. This explains why mathematical notation is often much less formal and precise than its users believe it to be \citep{BouteFunctionaldeclarativelanguage2005,SussmanRoleProgrammingFormulation2002}. An illustration is provided in Fig.~\ref{fig:scientific-discourse}, which shows a simple line of reasoning from elementary physics. Only a careful study of the text reveals that the parts typeset in blue correspond to formal reasoning. One way to identify these parts is to try to replace the textual description as much as possible by output from a computer algebra system. The parts typeset in black introduce the context (Newtonian physics) and define the formal symbols used in the equations in terms of physical concepts.

\begin{figure}[ht]
\centering
\fbox{
\begin{minipage}[c]{0.8\linewidth}
\textbf{Motion of a mass on a spring}

\vspace{1mm}
\small
We consider a point-like object of mass $m$ attached to a spring of force constant $k$ whose mass we assume to be negligible. The other end of the spring is attached to a wall. When the particle is at position $x$, the force acting on it is given by
\begin{equation}
  F = -k \cdot d,
\end{equation}
where $d = x - l$ is the displacement of $x$ relative to the spring's equilibrium length $l$. Newton's equation of motion for the mass takes the form
\begin{equation}
  F = m \ddot{x} = -k \cdot (x-l).
\end{equation}
{\color{blue}This second-order ordinary differential equation, which can be rewritten as
\begin{equation}
  \ddot{d} = -\frac{k}{m} d
\end{equation}
in terms of the displacement $d = x-l$, has the solution
\begin{equation}
  d(t) = A \cos(\omega t + \delta),
\end{equation}
where $\omega=k/m$} is the angular frequency of the oscillatory motion,
and the amplitude {\color{blue} $A$} and phase {\color{blue} $\delta$} are \textcolor{blue}{arbitrary real numbers.}
\end{minipage}
}
\caption{Mixing informal and formal reasoning in scientific discourse. The blue parts describe formal reasoning. The black parts establish the context and define the interpretation of the formal equations.}
\label{fig:scientific-discourse}
\end{figure}

Computers have vastly broadened the possibilities of formal reasoning through automation. Moreover, the fact that computation enforces a clear distinction of formal and informal reasoning makes it a useful intellectual tool in itself \citep{KnuthComputerScienceits1974,SussmanRoleProgrammingFormulation2002}. However, computing has led to a complete separation of automated formal reasoning from the informal narratives of scientific discourse. Even in the ideal case of a publication applying today's best practices for reproducible research, the reader has to figure out how text and mathematical formulas in the paper relate to the contents of the various files making up the executable computation.

This separation creates an important obstacle to verification. In the human-only scenario, both informal and formal reasoning are verified by a single person who, like the author, would not particularly care about the distinction. In the computer-assisted scenario, the narrative on its own cannot be verified because it is incomplete: the formal parts of the reasoning are missing. The computation on its own can be partially checked using software engineering techniques such as testing or static type checking, but in the absence of a specification, verification must remain incomplete. No amount of testing and verifying on the software side can verify that the computation actually does what it is expected to do. In terms of the illustration of Fig.~\ref{fig:scientific-discourse}, no verification restricted to the blue parts can establish that Eq.~(2) is the correct equation to solve, and no human examination of the black parts can establish that the computation correctly solves Eq.~(2).

To the best of my knowledge, the causes of mistakes in computer-aided research have never been analyzed systematically in a scientific study. However, my personal experience from 30~years of research in computational physics and chemistry suggests that roughly half of the mistakes that persist in spite of careful checks at all levels, in other words the mistakes that are detected after rather than before the publication of the results, can be described as ``the computation was perfectly reasonable but did not correspond to the scientific problem as described in the paper''. One typical manifestation of this problem is a mistake in a numerical constant in software source code, which includes the frequent case of a wrong sign. This was the cause for a widely publicized series of retractions of published protein structures, following the discovery of a sign error in data processing software \citep{MillerScientistNightmareSoftware2006}. Another variant is papers describing the computation only incompletely and in such a way that readers assume the computation to be different from what it actually was. A well-known example is the paper by Harvard economists Reinhart and Rogoff that supported austerity politics \citep{ReinhartGrowthTimeDebt2010}. It was based on an initially unpublished Excel spreadsheet whose later inspection by independent researchers revealed assumptions not mentioned in the paper, and mistakes in the implementation of some formulas \citep{HerndonDoeshighpublic2014}. A third variant, perhaps even the most frequent one, is scientists using software without knowing what it does exactly, leaving them unable to judge the well-foundedness of the methods that they are applying. A high-impact example concerning the analysis of fMRI brain scans was recently described \citep{EklundClusterfailureWhy2016}.

It is useful to look at this problem as a case of Human-Computer Interaction (HCI), the human part being the informal scientific discourse and the computer part being the computation. This point of view makes it clear that no improvement can be expected from focusing exclusively on scientific discourse or on software. In order to prevent the kinds of mistakes that I have described, scientists must have precise control over what software does when writing it, and a precise understanding of what software does when they take the user's or the verifier's role.

The popularity of computational notebooks, introduced in~1988 with the computer algebra system Mathematica \citep{WolframResearchIncMathematicaVersion1988} and later implemented for a wide range of programming languages by the Jupyter project \citep{KluyverJupyterNotebookspublishing2016}, shows that the re-unification of informal and formal reasoning into a single document corresponds to a real need in the scientific community. The computational notebook is a variant of the earlier idea of literal programming \citep{KnuthLiterateprogramming1984}, which differs in that the formal element embedded into the narrative is not a piece of software but a computation with specific inputs and outputs. Its popularity is even more remarkable in view of the restrictions that today's implementations impose: a notebook contains a single linear computation, allowing neither re-usable elements nor an adaptation of the order of presentation to the structure of the surrounding narrative. Both restrictions are a consequence of the underlying computational semantics: the code cells in a notebook are sent, one by one, as input to a command-line interpreter. Non-linear control flow can only happen inside a cell, and there is no way to refer to another cell, or sequence of cells, in order to re-use its contents. Notebooks can therefore capture only the top layer of the software stack shown in Fig.~\ref{fig:software-stack}, and even that only for relatively simple cases.

Another example for the re-unification of informal and automated formal reasoning is given by textbooks that include short computer programs to explain scientific concepts. Most of them deal specifically with computational science and use the code as examples, e.g. \citet{LangtangenPrimerScientificProgramming2012}, but some aim at conveying more fundamental scientific concepts using executable code for preciseness of notation \citep{SussmanFunctionaldifferentialgeometry2013,SussmanStructureinterpretationclassical2014}. Unfortunately, the code in such textbooks is in general not directly executable because they were prepared using traditional editing technology in view of being printed on paper. On the other hand, today's computational notebooks are not flexible enough to handle such more complex computational documents, which illustrates that the combination of narratives with computational content is still in its infancy.

\section{Human-computer interaction in computer-aided research}
\label{sec:hci}

Computer programs are predominantly viewed as tools, not unlike physical devices such as cars or microscopes. Humans interacting with tools can take on different roles. For software, the main roles are ``developer'' and ``user'', whereas for physical devices the number of roles tends to be larger: ``designer'', ``producer'', ``maintenance provider'', and ``user''. Software developers interact with software at the source code level using software development tools. Software users interact with the same software via a user interface designed by its developers. Some software provides several levels of user interfaces, for example by providing a scripting or extension programming language for an intermediate category of ``power users''. Each role in interacting with software is associated with a different mental model of how the software works. The basic user's mental model is restricted to \textit{what} the software does. A power user knows \textit{how} the software accomplishes its tasks, i.e. what the basic algorithms and data structures are. Developers also need to be aware of the software's architecture and of many implementation details.

In the development of scientific software, these different roles and the associated mental models have so far hardly been taken into account. In fact, in many scientific disciplines there are no clearly defined roles yet that people could adopt. More generally, human-computer interaction in computer-aided research has been shaped by historical accidents rather than by design. In particular, most software user interfaces are the result of a policy giving highest priority to rapid development and thus ease of implementation.

The analysis of verification that I have given in sections~\ref{sec:verification} and~\ref{sec:scientific-discourse} suggests that the users' minimal mental model of scientific software must include everything that may affect the results of a computation. This is a condition for scientists being able to verify the interface between informal and formal reasoning, i.e. to judge if a computation provides an answer to the scientific question being asked. Technical details can then be safely left to specialists. An important kind of specialist knowledge in many fields of scientific computing concerns performance characteristics such as the use of CPU time and memory. Performance experts would choose the best hardware and software for a given task, and work with developers on fine-tuning software. The extreme agility requirements in scientific research may well require many people to adopt multiple roles, such as user and performance expert, or performance expert and developer. However, verifying and working with the results should never require more than user-level knowledge.

In the rest of this article, I will concentrate on human-computer interaction at the user level, focusing on the interplay between informal and computer-assisted formal reasoning in scientific discourse. The overall goal is to explore how computations can be defined in such a way that human scientists can most easily understand and verify them.

\subsection{Case study: simulating the predator-prey equations}
\label{sec:example}

A simple example will help to illustrate the interface between informal and formal reasoning and between humans and computers. The predator-prey equations, also known as the Lotka-Volterra equations, describe the dynamics of the populations of two interacting species in an ecosystem in terms of non-linear differential equations. They take the form
\begin{eqnarray}
  \label{eq:lotka-volterra}
  \frac{dx}{dt} & = & \alpha x - \beta x y \\
  \frac{dy}{dt} & = & - \gamma y + \delta x y ,
\end{eqnarray}
where $x$ is the number of prey, $y$ the number of predators, and the positive constants $\alpha$, $\beta$, $\gamma$, and $\delta$ describe the events that change these numbers: $\alpha$ is the birth rate of prey, $\beta$ the rate at which prey are eaten by predators, $\gamma$ the death rate of predators, and $\delta$ the food-dependent birth rate of predators. These equations are based on a number of assumptions that a textbook or journal article --informal reasoning -- would discuss for each potential application.

For given parameters $\alpha$, $\beta$, $\gamma$, $\delta$, and given initial values $x(t_0), y(t_0)$, the predator-prey equations fully define $x(t)$ and $y(t)$ for all $t > t_0$, i.e. they are a complete specification for their solution. However, the equations do not provide an algorithm for actually finding a solution. To this day, no closed-form solution is known. Numerical solutions can be obtained after an approximation step that consists of discretization. For simplicity of presentation, I will use the simple Euler discretization scheme, even though it is \textit{not} a good choice in practice. This scheme approximates a first-order differential equation of the form
\begin{equation}
  \label{eq:ode}
  \frac{dz}{dt} = f(t, z(t))
\end{equation}
by the discretized equation
\begin{equation}
  \label{eq:finite-difference}
  z(t + h) = z(t) + h f(t, z(t)),
\end{equation}
which can be iterated, starting from $t=t_0$, to obtain $z(t)$ for a discrete set of time values $t + n h$ for any natural number $n$. For the predator-prey equations, $z(t)$ is replaced by the two components $x(t), y(t)$, which does not entail any fundamental change to the Euler method.

The discretized equation can be solved exactly using rational arithmetic. However, for performance reasons, rational arithmetic is usually approximated by inexact floating-point arithmetic. This approximation involves two steps:
\begin{enumerate}
  \item The choice of a floating-point representation with associated rules of arithemtic. The most popular choices are the single- and double-precision binary representations of IEEE standard 754-2008.
  \item The choice of the order of the floating-point arithmetic operations, which due to rounding errors do not respect the usual associativity rules for exact arithmetic.
\end{enumerate}

The scientific decomposition of this computation thus consists of five parts, each of which requires a justification or discussion in an informal narrative:
\begin{enumerate}[label=(\Alph*)]
  \item The description of the scientific question by the predator-prey equations.
  \item The values of the constant parameters.
  \item The values of the initial values $x(t_0)$ and $y(t_0)$.
  \item The discretization using the Euler method and the choice of $h$.
  \item The choices concerning the floating-point approximation: precision, rounding mode, order of arithmetic operations.
\end{enumerate}

The technical decomposition of a typical implementation of this computation looks very different:
\begin{enumerate}
  \item A program that implements an algorithm derived from the predator-prey equations, using partially specified floating-point arithmetic. This program also reads numerical parameters from a file, and calls a function from an ODE solver library.
  \item An ODE solver library implementing the Euler method in partially specified floating-point arithmetic.
  \item An input file for the program that provides the numerical values of the parameters, the initial values $x(t_0)$ and $y(t_0)$, and the step size $h$.
  \item A compiler defining the precise choices for floating-point arithmetic.
\end{enumerate}

The transitions from A to 1, from B/C/D to 3, and from D/E to 1/2/4 require human verification because they represent transitions from informal to formal reasoning. The two approximations that require scientific validation are A$\to$D (the Euler method) and D$\to$E (floating-point approximation). In this validation, formal reasoning (running the code) is an important tool. The first validation is in practice done empirically, by varying the step size $h$ and checking for convergence. The many subtleties of this procedure are the subject of numerical analysis. The validity of the floating-point approximation would be straightforward to check if the computation could be done in exact rational arithmetic for comparison. This is unfortunately not possible using the languages and libraries commonly used for numerical work, which either provide no exact rational arithmetic at all or require the implementation 1/2 to be modified, introducing another opportunity for mistakes.

% In addition to these rather generic verification and validation approaches that can be applied to most differential equations, there are methods specific to a particular set of equations. As an example for the case of the predator-prey equations, the solutions are periodic and equations can be derived for the closed curves describing $y(t)$ vs. $x(t)$. The numerical solutions can be compared to these equations.

\section{Digital Scientific Notations}
\label{sec:dsn}

A \textit{digital scientific notation} is a formal language that is part of the user interface of scientific software. This definition includes, but is not limited to, formal languages embedded into informal scientific discourse, as in the case of computational notebooks. Another important category contains the file formats used to store scientific datasets of all kinds. As I have explained in section~\ref{sec:hci}, the scientific information to be expressed in terms of digital scientific notations includes everything that is relevant at the user level, i.e. everything that has an impact on the results of a computation. This includes in particular scientific models and computational methods, but also more traditional datasets containing, for example, experimental observations, simulation results, or fitted parameters.

Digital scientific notations differ in two major ways from traditional mathematical notation:
\begin{enumerate}
\item They must be able to express algorithms, which take an ever more important role in scientific models and methods.
\item They must be adapted to the much larger size and much more complex structure of scientific models and data that can be processed with the help of computers.
\end{enumerate}

The first criterion has led to the adoption of general-purpose programming languages as digital scientific notations. Most published computational notebooks, for example, use either Python or R for the computational parts. The main advantage of using popular programming languages in computational documents is good support in terms of tools and libraries. On the other hand, since these are programming rather than specification languages, they force users to worry about many technical details that are irrelevant to scientific knowledge. For example, the Python language has three distinct data types for sequences: lists, tuples, and arrays. There are good technical reasons for having these separate types, but for scientific communication, the distinction between them and the conversions that become necessary are only a burden. Another disadvantage is that the source code of a programming language reduces all scientific knowledge to algorithms for computing specific results. This process implies a loss of information. For example, the predator-prey equations from Eq.~\ref{eq:lotka-volterra} contain information that is lost in a floating-point implementation of its discrete approximation because the latter can no longer be used to deduce general properties of exact solutions.

Domain-specific languages (DSLs) are another increasingly popular choice for representing scientific knowledge. In contrast to general-purpose programming languages, DSLs are specifically designed as digital scientific notations, and usually avoid the two main disadvantages of programming languages mentioned above. Most scientific DSLs are embedded in a general-purpose programming language. A few almost arbitrarily selected examples are Liszt, a DSL for mesh definitions in finite-element computation, embedded in Scala \citep{DeVitoLisztdomainspecific2011}, Kendrick, a DSL for ODE-based models in epidemiology, embedded in Smalltalk \citep{BuiSeparationconcernsepidemiological2016}, and an unnamed DSL for micromagnetics, embedded in Python \citep{BegUserinterfacescomputational2017}. The choice for an embedded DSL is typically motivated by simpler implementation and integration into an existing ecosystem of development tools and libraries. On the other hand, embedded DSLs are almost impossible to re-implement in a different programming language with reasonable effort, which creates a barrier to re-using the scientific knowledge encoded using them. A stand-alone DSL is independent of any programming language, as is illustrated by Modelica \citep{FritzsonModelicaunifiedobjectoriented1998}, a general modeling language for the natural and engineering sciences for which multiple implementations in different languages exist. However, each of these implementations is a rather complex piece of software.

Looking at how scientific DSLs are used in practice, it turns out that both embedded and stand-alone DSLs end up being a user interface for a single software package, or at best a very small number of packages. Adopting an existing DSL for a new piece of software is very difficult. One obstacle is that the existing DSLs can be too restrictive, having been designed for a narrowly defined domain. For embedded DSLs, interfacing the embedding language with the implementation language of the new software can turn out to be a major obstacle. Finally, the complexity of a DSL can be prohibitive, as in the case of Modelica. In all these scenarios, the net result is a balkanization of digital scientific knowledge because for each new piece of software, designing a new DSL is often the choice of least effort.

These considerations lead to two important criteria for good digital scientific notations that existing ones do not satisfy at the same time: 
\begin{itemize}
\item Generality. While it is unrealistic to expect that a single formal language could be adequate for computer-aided research in all scientific disciplines, it should be usable across narrowly defined domains of research, and be extendable to treat newly discovered scenarios.

\item Simplicity. The implementation of user interfaces based on a digital scientific notation should not require a disproportionate effort compared to the implementation of the scientific functionality of a piece of software.
\end{itemize}

In the following, I will describe an experimental digital scientific notation that was designed with these criteria in mind, and report on first experiences with simple toy applications. While it is too early to judge if this particular notation will turn out to be suitable for real-life applications, it illustrates that better digital scientific notations can be designed if their role at the human-computer interface is fully taken into account.

\section{Leibniz, a digital scientific notation for continuous mathematics}
\label{sec:leibniz}

An important foundation of many scientific theories is the mathematics of smoothly varying objects such as the real numbers. This foundation includes in particular geometry, analysis, and linear algebra. In some scientific disciplines, such as physics and chemistry, this is the dominant mathematical foundation. In other disciplines, such as biology, it is one important foundation among others, notably discrete mathematics. The digital scientific notation \bluehref{https://github.com/khinsen/leibniz}{Leibniz}, named after 17th-century polymath Gottfried Wilhelm Leibniz, focuses on continuous mathematics and its application. Like many computer algebra systems, but unlike common programming languages, it can express functions of real numbers and equations involving such functions, in addition to the discrete approximations using rational or floating-point numbers that are used in numerical work.

The design priorities for Leibniz are:
\begin{itemize}
\item Embedding in narratives such as journal articles, textbooks, or software documentation, in order to act as an effective human-computer interface. The code structure is subordinate to the structure of the narrative.
\item Generality and simplicity, as discussed in section~\ref{sec:dsn}.
\end{itemize}

Before discussing how Leibniz achieves these goals, I will present the language through a few illustrative examples.

\begin{figure}[ht]
\centering
\fbox{
\begin{minipage}[c]{0.7\linewidth}
\includegraphics[width=\linewidth]{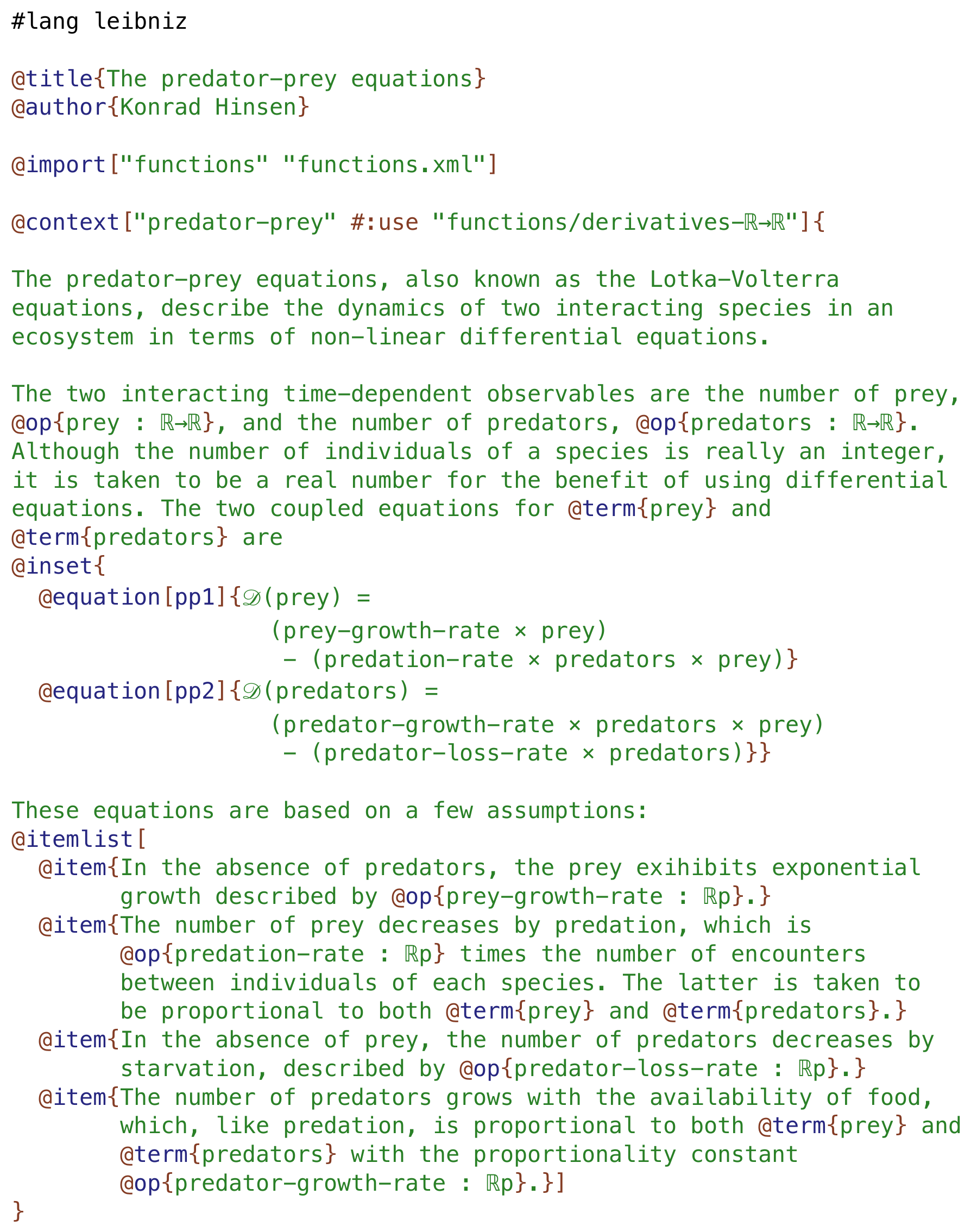}
\end{minipage}
}
\caption{The author view of a Leibniz document shows Leibniz code embedded in a narrative. Most of the commands (starting with \texttt{{@}}) are inherited from the Scribble document language, only \texttt{{@}context}, \texttt{{@}op}, \texttt{{@}term}, and \texttt{{@}equation} are added by Leibniz. This example defines a single context called \texttt{predator-prey} that uses another context called \texttt{derivatives-{\small$\mathbb{R}$}$\rightarrow${\small$\mathbb{R}$}} that is defined in an imported Leibniz document \texttt{functions}.
}
\label{fig:leibniz-source}
\end{figure}

\begin{figure}[ht]
\centering
\fbox{
\includegraphics[width=0.9\linewidth]{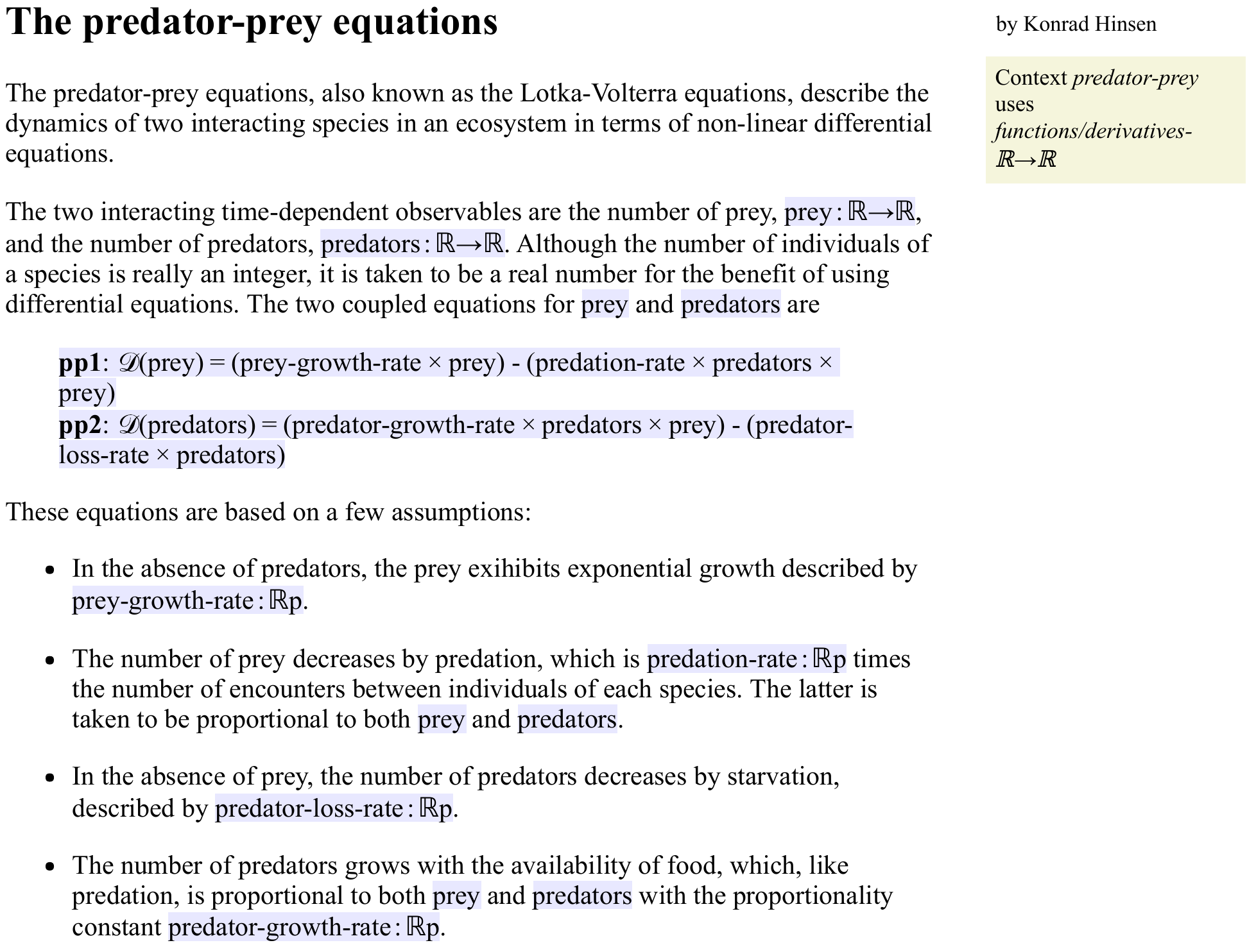}
}
\caption{The reader view of a Leibniz document. All code is shown on a blue background.}
\label{fig:leibniz-rendered}
\end{figure}

\begin{figure}[ht]
\centering
\fbox{
\includegraphics[scale=0.5]{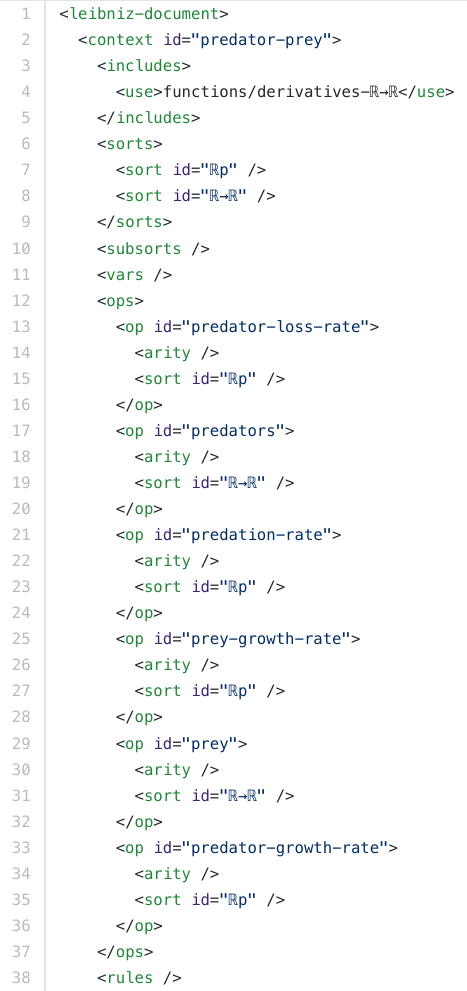}
}
\fbox{
\includegraphics[scale=0.5]{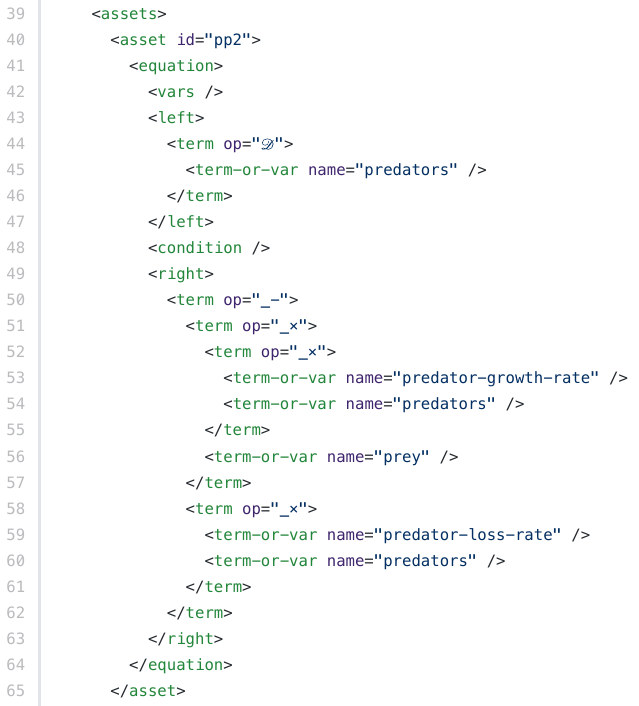}
}
\fbox{
\includegraphics[scale=0.5]{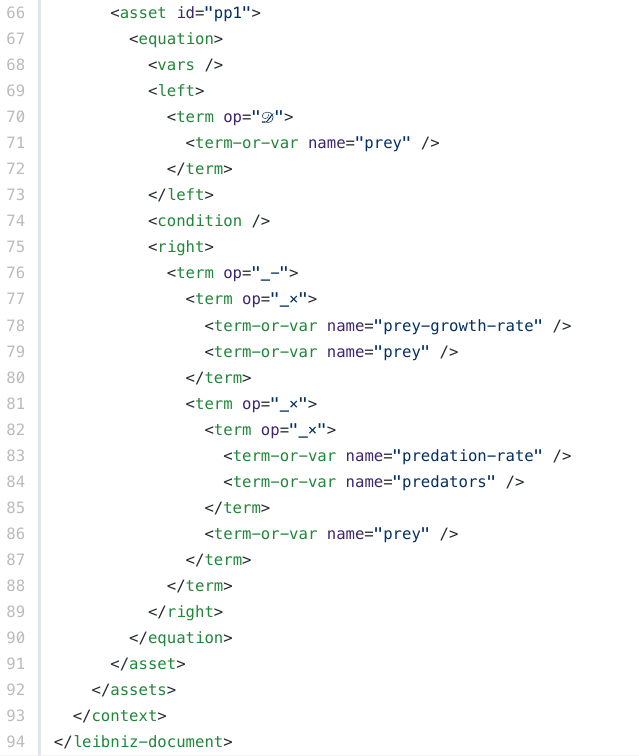}
}
\caption{The machine-readable view of the predator-prey example.}
\label{fig:leibniz-xml}
\end{figure}

Figs.~\ref{fig:leibniz-source}, \ref{fig:leibniz-rendered}, and~\ref{fig:leibniz-xml} show three views of a Leibniz document introducing the predator-prey equations from section~\ref{sec:example}. This and other examples are also available on-line in the \bluehref{http://khinsen.net/leibniz-examples/}{Leibniz example collection}. Fig.~\ref{fig:leibniz-source} shows the author view. Leibniz is implemented as an extension to the document language Scribble \citep{FlattScribbleClosingbook2009}, which is part of the Racket ecosystem \citep{FelleisenRacketManifesto2015}. Scribble source code is a mixture of plain text and commands, much like the better known document language \LaTeX\  \citep{LamportLATEXdocumentpreparation1994}. Commands start with an \texttt{{@}} character. Leibniz adds several commands such as \texttt{{@}op} and \texttt{{@}equation}, which define elements of Leibniz code. The Leibniz processing tool generates the two other views from the author's input document. Fig.~\ref{fig:leibniz-rendered} shows the reader view, a rendered HTML page, in which the Leibniz code is typeset on a blue background. This makes the transition between informal and formal reasoning visible at a glance. The machine-readable view, shown in Fig.~\ref{fig:leibniz-xml}, is an XML file that represents the code in a very rigid format to facilitate processing by scientific software.

% Since the Leibniz language is likely to undergo many changes in the future, I will not give a detailed description here. Instead, I will concentrate on those features that are important for a digital scientific notation embedded into informal scientific discourse.

In terms of semantics, Leibniz' main source of inspiration has been the OBJ family of algebraic specification languages \citep{GoguenIntroducingOBJ2000}, and in particular its most recent incarnation, Maude  \citep{ClavelMaudeSpecificationprogramming2002}. In fact, the semantics of the current version of Leibniz are a subset of Maude's functional modules, the main missing features being conditional sort membership and the possibility to declare operators as commutative and/or associative. As I will discuss later, this minimalist language is not sufficient for dealing with the complex scientific models used in real research. Various features will be added in the future as the need becomes evident in practical applications.

\begin{figure}[ht]
\centering
\fbox{
\includegraphics[width=0.9\linewidth]{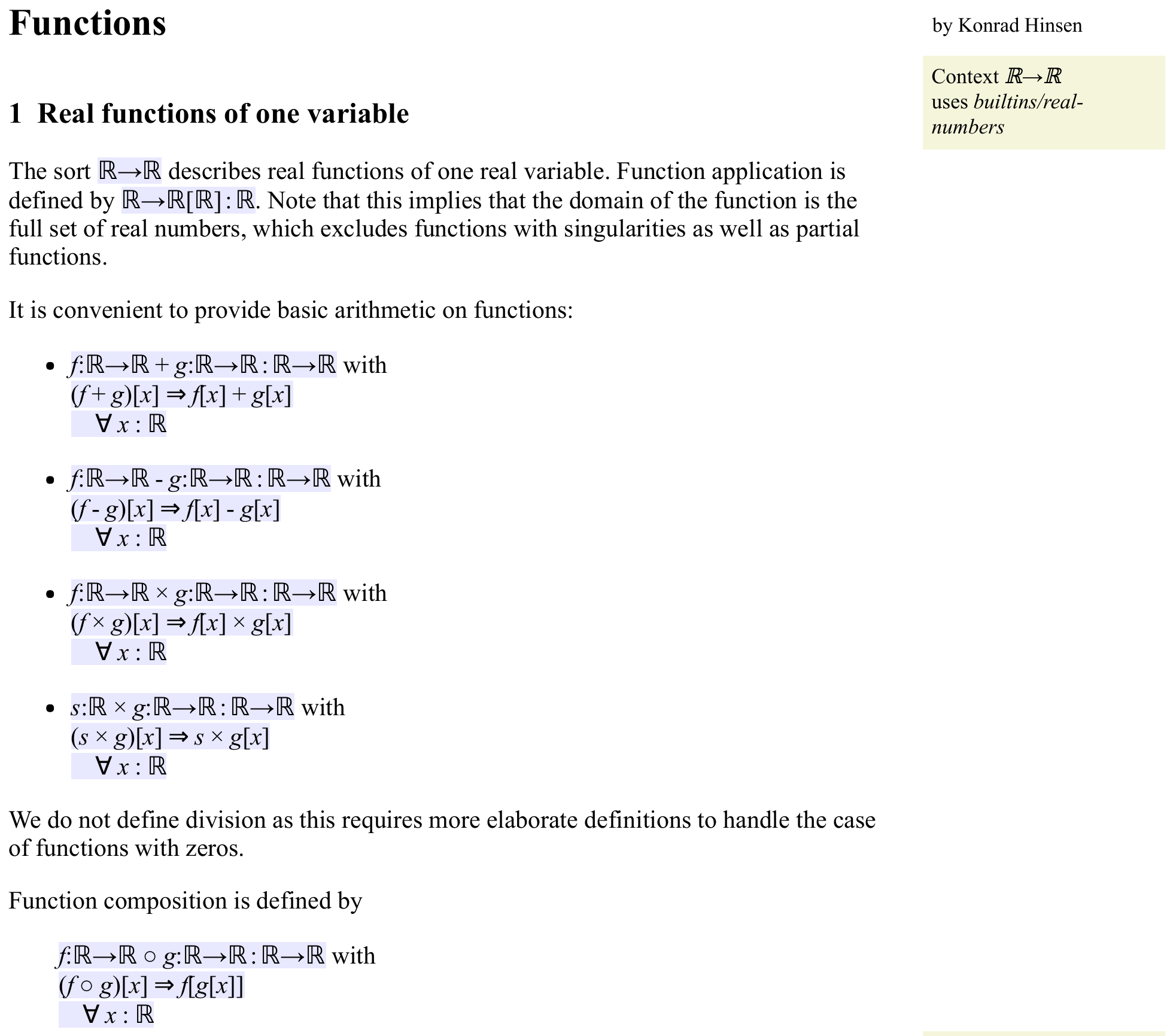}
}
\caption{The beginning of the document \texttt{functions} referred to by the predator-prey example. The full version is \bluehref{http://khinsen.net/leibniz-examples/examples/functions.html}{available online}.}
\label{fig:leibniz-functions}
\end{figure}

As the reference to the OBJ family suggests, Leibniz is based on term rewriting. The code units in Leibniz are called \textit{contexts} (corresponding essentially to Maude's functional modules) and consist of (1) the definition of an order-sorted term algebra, (2) a list of rewrite rules, and (3) any number of assets, which are arbitrary values (terms or equations) identified by unique labels. A Leibniz document, such as the one shown in Figs.~\ref{fig:leibniz-source}, \ref{fig:leibniz-rendered}, and~\ref{fig:leibniz-xml}, is a sequence of such contexts, each of which is identified by a unique name. A context can \textit{use} another context, inheriting its term algebra and its rewrite rules, or \textit{extend} it, in which case it also inherits its variables and assets. In a typical Leibniz document, each context extends the preceding one, adding or specializing scientific concepts. This corresponds to a frequent pattern of informal reasoning in scientific discourse that starts with general concepts and assumptions and then moves on to more specific ones. The ``use'' relation typically serves for references to contexts imported from other documents that treat a more fundamental theory or methodology. In the example, the context \texttt{predator-prey} uses a context from another document (shown partially in Fig.~\ref{fig:leibniz-functions} and \bluehref{http://khinsen.net/leibniz-examples/examples/functions.html}{available online}) called \texttt{functions} that defines real functions of a single real variable having derivatives. When using or extending contexts, it is possible to specify transformations, in particular renaming to avoid name clashes. A small number of builtin contexts defines booleans and a hierarchy of number types with associated arithmetic operations.

Like in the OBJ family, a term algebra is defined by operator declarations, which in turn refer to sorts defined by sort declarations. All these declarations can be inserted in arbitrary order into the narrative contained in the body of a \texttt{{@}context} declaration, and can also be repeated. The predator-prey example contains no sort or subsort declarations of its own, but it adds six nullary operators, representing the predator and prey populations and the four rate constants, to the term algebra it inherits from \texttt{functions/derivatives-{\small$\mathbb{R}$}$\rightarrow${\small$\mathbb{R}$}}. The sort \texttt{{\small$\mathbb{R}$}} stands for the real numbers, the sort \texttt{{\small$\mathbb{R}$}p} for the positive real number, and the sort \texttt{{\small$\mathbb{R}$}$\rightarrow${\small$\mathbb{R}$}} for real functions of one real variable. Other than these nullary operators, the context \texttt{predator-prey} only defines two assets, the two differential equations for the predator and prey populations, identified by the labels \texttt{pp1} and \texttt{pp2}. An example for binary infix operators with rewrite rules can be found in the context \texttt{functions/derivatives-{\small$\mathbb{R}$}$\rightarrow${\small$\mathbb{R}$}} partially shown in Fig.~\ref{fig:leibniz-functions}. It defines rules for the sum, difference, and product of real functions. Contrary to sort and operator declarations, the order of rewrite rules is important because when multiple rules are eligible for rewriting a term, the textually first one is selected.

The context \texttt{functions/derivatives-{\small$\mathbb{R}$}$\rightarrow${\small$\mathbb{R}$}} also illustrates one of the few special operators in Leibniz, the bracket operator, which is merely syntactic sugar, the term \texttt{a[b]} being treated like \texttt{[](a, b)}, except that \texttt{[]} is not legal operator syntax. The two other special operators are superscript and subscript: $a_b$ is equivalent to \texttt{\_(a, b)} and $a^b$ becomes \texttt{\^(a, b)}. In addition to these three special operators, contexts can define arbitrary prefix operators of the form \texttt{op(a, b)}, with any number of arguments, and arbitrary binary infix operators of the form \texttt{a op b}. There are no precedence rules for infix operators in Leibniz, the use of round brackets is obligatory to resolve ambiguities. The sole exception is a chain of identical binary operators at the same level of the expression. For example, \texttt{a + b + c} is allowed and equivalent to \texttt{(a + b) + c}. This rule has been inspired by the Pyret language \citep{ProgrammingPyret2018} and is a compromise between the familiarity of the precedence rules in mathematics and the ease of not having to remember precedence values for a potentially large number of infix operators. The current Leibniz examples also make extensive use of mathematical Unicode symbols in order to define familiar-looking operators. This is, however, a question of style rather than a language feature.

Another \bluehref{http://khinsen.net/leibniz-examples/examples/heron.html}{online example} illustrates how Leibniz can help to document and validate approximations. The example implements Heron's algorithm for computing square roots, which is a special case of the Newton-Raphson method for finding the roots of a function. The algorithm is first developed for real numbers, with test computations using exact rational number arithmetic. A conversion tool that is part of the Leibniz implementation then derives a floating-point version of the algorithm using the 64-bit binary representation of IEEE standard 754-2008, keeping the order of arithmetic operations from the original algorithm, which is unambiguous in Leibniz. This example also showcases another user interface feature: in the reader view, computationally derived information is typeset on a green background, making it easy to distinguish from the human input typeset on a blue background.

A final feature of Leibniz that deserves discussion is its type system, or rather sort system as it is habitually called in the context of formal logic and term algebras. In this system, directly taken over from Maude, sort and subsort declarations define a directed acyclic sort graph, which in general consists of multiple connected components called kinds. Operator declarations assign a sort to each term and a required sort to each argument position of a non-nullary operator. Mismatches at the kind level, i.e. an argument sort not being in the same connected component as a required sort, lead to a rejection of a term in what resembles static type checking in programming languages. Mismatches inside a kind, however, are tolerated. The resulting term is flagged as potentially erroneous but can be processed normally by rewriting. If in the course of rewriting the argument gets replaced by a value that is a subsort of the required type, the error flag is removed again. The presence of an error flag on a result of a computation thus resembles a runtime error in a dynamically typed language. This mixed static-dynamic verification system offers many of the benefits of a static type checker, but also allows the formulation of constraints on values that cannot be verified statically. Leibniz uses this feature to define fine-grained subsorts on the number sorts, e.g. ``positive real number'' or ``non-zero rational number'', which turn out to be very useful in many scientific models.

\section{Discussion}

The three main goals in the development of Leibniz are (1) its usability as a digital scientific notation embedded in informal narratives, (2) generality in not being restricted to a narrowly defined scientific domain, and (3) simplicity of implementation in scientific software. While the current state of Leibniz, and in particular the small number of test applications that have been tried, do not permit a final judgment on how well these goals were achieved, it is nevertheless instructive to analyze \textit{which} features of Leibniz are favorable to reaching these goals and how Leibniz compares to the earlier digital scientific notations reviewed in section~\ref{sec:dsn}.

The key feature for embedding is the highly declarative nature of Leibniz. The declarations that define a context and the values that build on them (terms and equations) can be inserted in arbitrary order into the sentences of a narrative. Verification at the informal-formal borderline is as well supported by Leibniz as by traditional mathematical notation. None of the digital scientific notations in use today shares this feature. Order matters only for rewrite rules, which has not appeared to be a limitation in the experiments conducted so far. Leibniz permits to write rules as assets identified by unique labels, and then assemble a list of named assets into a rule set for rewriting, but so far this feature has not found a good use.

Visual highlighting of the formal parts of a narrative (the blue and green background colors) allows readers to spot easily which parts of a narrative can affect a computation. Moreover, the reader can be assured that the internal coherence of all such highlighted information has been verified by the Leibniz authoring tool. For example, an equation typeset on a blue background is guaranteed to use only operators declared in the context and the sorts of all terms have been checked for conformity. In this way, Leibniz actively supports human verification.

Generality is achieved by Leibniz not containing any scientific information and yet encapsulating useful foundations for expressing it. These foundations are term algebras, equational logic, and numbers as built-in terms. This is an important difference in comparison to scientific DSLs. In fact, the analogue of a scientific DSL is not Leibniz, but a set of domain-specific Leibniz contexts. The common foundation makes it possible to combine contexts from different domains, which is difficult with DSLs designed independently. General-purpose programming language follow the same approach as Leibniz in providing domain-neutral semantic foundations for implementing algorithms. These foundations are usually lambda calculus, algebraic data types, and a handful of built-in basic data types such as numbers and character strings. Leibniz' main advantage is in this respect is that equational logic is a more useful foundation for scientific knowledge than lambda calculus.

The principle of factoring out application-independent structure and functionality has a practically successful precedent in data languages such as XML \citep{BrayExtensibleMarkupLanguage2006}. The foundation of XML is a versatile data structure: a tree whose nodes can have arbitrary labels. XML defines nothing but the syntax for this data structure, delegating the domain-specific semantics to schemas. The combination of data referring to different schemas is made possible by the XML namespace mechanism. The machine-facing side of Leibniz can be thought of as a layer in between the pure syntax of XML and domain-specific scientific knowledge, providing a semantic foundation for scientific models and methods.

The separation of syntax and semantics in XML is reflected by tools that process information stored in XML-based formats. Domain-specific tools can delegate parsing and a part of validation to generic parsers and schema validators. This same principle is expected to ensure simplicity of implementation for Leibniz. Validating and rewriting terms are generic tasks that can be handled by a domain-independent Leibniz runtime library. Assuming Leibniz is widely adopted, optimized Leibniz runtimes will become as ubiquitous as XML parsers.

Another aspect of Leibniz that makes is easier to use in scientific software is its clear separation of a machine-facing syntax based on XML from the representations that human users interact with. This is in striking contrast to general-purpose programming languages and stand-alone DSLs, whose syntax \textit{is} their user interface. This makes it difficult to extract and analyze information stored in a program, because any tool wishing to process the source code must deal with the non-trivial syntax designed for human convenience. Moreover, suitable parsers are rarely available as reusable libraries.

\section{Outlook}

The work reported in this article focused on the human-computer interaction aspect of digital scientific notations. The initial semantics of Leibniz were somewhat arbitrarily chosen to be a subset of Maude, which looked like a good starting point for first experiments, some of which have been discussed above. They are encouraging as to the principles of Leibniz' design, in particular concerning embedding in informal scientific discourse. However, these first experiments also suggest that the language is currently too minimalist for productive use in computer-aided research. In particular, the lack of predefined collections (lists, sets) makes it cumbersome to deal with many common situations. Another aspect of the language that deserves further attention is the sort system summarized in section~\ref{sec:leibniz}. Many common value constraints in scientific applications would require value-dependent sorts, in the spirit of dependent types. Examples are the compatibility of units of measure, or of the dimensions of matrices.

Another aspect that will require further work is the definition of the role of digital scientific notations such as Leibniz in the ecosystem of scientific software. A theoretically attractive but at this time not very feasible approach would have software tools read specifications from Leibniz documents and perform the corresponding computations. In addition to the fundamental issue that we do not know yet how to turn specifications into efficient implementations automatically, there is the practical issue that today's scientific computing universe is very tool-centric, with users often adapting their research methodology to their tools rather than the inverse. A more realistic short-term scenario sees Leibniz used in the documentation of software packages, which could then contain a mixed informal/formal specification of the software's functionality. This specification could be verified scientifically by human readers, and the software could be verified against it using  techniques such as testing or formal verification. A scientific study would be documented in another Leibniz document that uses contexts from the software's specification.

Leibniz and digital scientific notations similar to it are also a promising candidates for unifying symbolic and numerical computation in science. As the example of the predator-prey equations shows, Leibniz can represent not only computations, but also equations. A computer algebra system could process equations formulated in Leibniz, producing results such as analytical solutions or approximations which could be expressed in Leibniz as well. Corresponding Leibniz contexts could be derived automatically from the OpenMath standard \citep{OpenMathsocietyOpenMath2000}.

Finally, there is obviously a lot of room for improvement in the tools used by authors and readers for interacting with Leibniz content. Ideally, the author and reader would work with identical or very similar views, which should be more interactive than plain text or HTML documents. Much inspiration, and probably also implementation techniques, can be adopted from computational notebooks and other innovations in scientific publishing that are currently under development.

\section*{Acknowledgments}

I am grateful to Prof. Shriram Krishnamurthi for recommending the adoption of the infix operator rules from the Pyret language, and for suggesting the predator-prey equations as an example for demonstration.

\bibliographystyle{plain}
\bibliography{preprint}

\end{document}